# Impact of Usability Mechanisms: A Family of Experiments on Efficiency, Effectiveness and User Satisfaction

Juan M. Ferreira, Francy Rodríguez, Adrián Santos, Silvia T. Acuña, Natalia Juristo

**Abstract**— **Context**: The usability software quality attribute aims to improve system user performance. In a previous study, we found evidence of the impact of a set of usability characteristics from the viewpoint of users in terms of efficiency, effectiveness and satisfaction. However, the impact level appears to depend on the usability feature and suggest priorities with respect to their implementation depending on how they promote user performance. **Objectives**: We use a family of three experiments to increase the precision and generalization of the results in the baseline experiment and provide findings on the impact on user performance of the Abort Operation, Progress Feedback and Preferences usability mechanisms. **Method**: We conduct two replications of the baseline experiment in academic settings. We analyse the data of 367 experimental subjects and apply aggregation (meta-analysis) procedures. **Results**: We find that the Abort Operation and Preferences usability mechanisms appear to improve system usability a great deal with respect to efficiency, effectiveness and user satisfaction. **Conclusions**: We find that the family of experiments further corroborates the results of the baseline experiment. Most of the results are statistically significant, and, because of the large number of experimental subjects, the evidence that we gathered in the replications is sufficient to outweigh other experiments.

**Index Terms**— Usability mechanism, efficiency, effectiveness, satisfaction, experimental software engineering, family of experiments.

—————————— ◆ ——————————

## 1. INTRODUCTION

USABILITY is a quality characteristic of a software system, which plays a more important role in highly interactive systems [1]–[3]. According to ISO/IEC 25010 [4], usability is defined as "the degree to which a product or system can be used by specified users to achieve specific goals with effectiveness, efficiency and satisfaction in a specified context of use". From the viewpoint of human-computer interaction (HCI), usability is related primarily to user interface design and user-system interaction [5]. However, there is evidence that some of these recommendations also affect system functionality and not only its interface [6]. HCI researchers propose recommendations for achieving a proper usability level in software systems [7]–[12].

Software engineering (SE) studies how to include these HCI recommendations in software development [13], and SE experimentation aims to find empirical evidence on both final system usability and how to implement and improve usability during the software development process. There are many studies on usability evaluation related to recommendations that affect graphical interface issues, but there are very few empirical studies that address usability recommendations that affect software design and measure their benefits from the viewpoint of users [14].

In order to extend empirical evidence on the impact of HCI recommendations that affect software design, the results of an experiment studying the effect on efficiency, effectiveness and user satisfaction of three usability mechanisms —Abort Operation (ABR), Progress Feedback (PFB) and Preferences (PRF)— was reported in [14]. Usability mechanisms are functionalities that should, according to the HCI community, be implemented within a software system to increase its usability. We have conducted two replications of this baseline experiment to build a family of three experiments. This paper illustrates how the results evolve from the baseline to the family of experiments.

More and more replications of experiments are being conducted in SE [15]. Different authors have analysed the process of experiment replication [16] and data aggregation techniques [17] in order to identify the best techniques for use in the field of SE. Moreover, there is unanimous agreement within the scientific community that one-off experiments are, with few exceptions, of little value. The truth is that the results of a baseline experiment can only be confirmed through replication and results comparison.

A family of experiments is a set of experimental replications with access to the raw (or aggregated) data of each of at least three experiments with at least two different technologies testing the same response variable according a known experimental design and protocol [17].

————————————————
- JM. Ferreira is with the Facultad Politécnica, Universidad Nacional de Asunción, CC 2111, San Lorenzo, Paraguay, e-mail: jferreira@pol.una.py.
- F. Rodríguez is with Universidad Católica de Ávila, Calle de los Canteros S/N, Ávila 05002, Spain, e-mail: fdiomar.rodriguez@ucavila.es.
- Santos is with the M3S (M-Group), ITEE University of Oulu, P.O. Box 3000, 90014, Oulu, Finland, e-mail: adrian.santos.parrilla@oulu.fi.
- ST. Acuña is with the Universidad Autónoma de Madrid, Calle Francisco Tomás y Valiente 11, 28049 Madrid, Spain, e-mail: silvia.acunna@uam.es.
- N. Juristo is with the Escuela Técnica Superior de Ingenieros Informáticos, Universidad Politécnica de Madrid, Campus Montegancedo, 28660 Boadilla del Monte, Spain, e-mail: natalia@fi.upm.es.

The aim of replication is to provide a family of experiments to aggregate separate experiments and get more reliable results, as well as to be able to analyse aspects that individual experiments have overlooked, providing accurate information for decision making and/or more in-depth knowledge of the issue under investigation. The aim of replications is to validate and round out the results of the baseline experiment [18].

In this study, the goal of the baseline experiment was to evaluate the impact of three usability mechanisms, ABR, PFB and PRF, on an online shop web application. The evaluation was carried out using three response variables taken from the usability definition set out in ISO/IEC 25010 [4]: efficiency, effectiveness and user satisfaction. The baseline experiment was conducted in two contexts — an academic and a non-academic setting— with 168 users divided into 24 experimental groups. Each group performs three online shopping tasks. Efficiency was measured using number of clicks and time taken, effectiveness was gauged by percentage task completion and user satisfaction was gathered from a questionnaire. There were 100 experimental subjects in Replication 1 and 99 in Replication 2, amounting to a total number of 367 subjects for the family of experiments.

The results of the baseline experiments [14] showed that the adoption of ABR has a significantly positive effect on efficiency, effectiveness and user satisfaction, the adoption of PFB does not appear to have any impact on any of the variables and the adoption of PRF has a significantly positive effect on effectiveness and user satisfaction, but no impact on efficiency. In no case do the usability mechanisms degrade user performance.

The replications were as similar as possible to the baseline experiment. Strict replications increase sample size and thus statistical power [19], [20]. The three experiments have the same between-subjects experimental design, the same goal, the same research hypotheses and the same two-level factors: adopted and not adopted. Replication 1 evaluates two of the three baseline experiment response variables: efficiency and satisfaction. Effectiveness data are missing from Replication 1. Replication 2 evaluates the same three response variables as the baseline experiment.

The **major contribution** of this paper is *that it provides evidence of how the three HCI recommendations improve the usability of a system using a family of experiments.* Data from three different samples can be aggregated by the family of experiments, leading to several findings.

> *Key findings*
> - The baseline experiment finding that Abort Operation significantly improves efficiency, effectiveness and user satisfaction is confirmed.
> - The results corroborate the fact that PFB has a negligible impact compared with the other two mechanisms, even though it is the costliest to implement
> - The family of experiments reveals that, like ABR, PRF also has a positive effect on efficiency, effectiveness and satisfaction. This contradicts the baseline experiment finding suggesting that PRF did not improve user efficiency (speed and interactivity).
> - The aggregated data of the family of experiments again suggest that the three mechanisms improve system usability and does not undermine user performance.

Studies using families of experiments have been conducted to explore SE issues like requirements analysis [21], the implications of the use of model-driven development (MDD) [20], [22]–[24] or how test-driven development (TDD) improves software quality [23], [25]. We also found a study aimed at empirically validating a web usability evaluation process that can be integrated into web development processes that use MDD [26]. However, we have not found any study using families of experiments to analyse the application of HCI recommendations to improve usability. This study aims to fill the gap in the literature.

**Paper Organization**. Section 2 describes work related to this research. Section 3 shows the design of the baseline experiment. Section 4 describes the replications and the results of their analysis. Section 5 analyses the results aggregation and discusses the results. Section 6 describes the internal and external, and statistical conclusion validity threats. Finally, Section 7 presents the conclusions and future works.

## 2. RELATED WORK

Even though usability is recognized as a software product quality feature [4], [27], many systems still do not achieve an acceptable level of usability [5], [28], [29]. The SE experimentation community has been studying usability from different viewpoints [14]. Some studies focus on the software process and lifecycle activities, whereas others focus on the end products. The former study how to implement or evaluate usability characteristics in the different software development lifecycle activities [6], [30], [31]. The latter focus on validating the usability of products, technologies and applications [32]–[35]. As found in [14], there are very few empirical studies that address usability recommendations that affect software design and measure their benefits from the viewpoint of users.

Within the empirical studies that evaluate or validate applications or final products, some focus on web applications [36]–[38], others on mobile applications [39], [40] and other evaluate specific properties like security [41], comprehension and learnability [42] or application programming interfaces (APIs) [43]. However, these empirical studies are mostly confined to one experiment. There are few studies that include replications. Although SE experimentation has increased over recent years [44], it still has the pitfall of using sample sizes that are too small to be representative [19], [45]. To overcome this shortcoming, researchers have resorted to experiment replication, which, through data aggregation, provides more evidence and increases the quality of the findings based on more evidence [44].

Separate experiments provide useful data for generating empirical evidence and improving existing knowledge,

but a larger sample size can lead to new discoveries that are not observed when running a single experiment [20]. Statistical methods perform efficiently as samples are larger [17]. During data aggregation, the effect sizes of all the replications are calculated first based on descriptive statistics, like means, variances or sample sizes or results of the experiment statistical tests and are then combined using a meta-analysis model [19], [46].

Techniques like aggregated data (AD), narrative synthesis, individual participant data stratified (IPD-S) and aggregation of p-values are used to analyse families of experiments. IPD-S and AD were found to be the best techniques for analysing families of SE experiments [17], and it has been established that all the data of the experiments that are part of the family have to be analysed jointly, recognizing their source experiment [19]. Fixed-effects and random-effects models can be fitted. For fixed-effects models, two-factor ―experiment and treatment― linear regression is usually used. For random-effects models, a mixed linear model, again with two factors ―experiment and treatment― is usually fitted.

Taking into account the importance of families of experiments and research on the best techniques to analyse results, we report this study on a family of experiments generated by two replications of the baseline experiment reported in [14]. The experiment addresses usability recommendations with a high impact on software design. We evaluate three quality attributes that, according to ISO/IEC 25010 [4], are useful for determining product usability: efficiency, effectiveness and satisfaction. An increase in these three quality characteristics is a measure of impact on usability, which can improve or degrade application usability. The experiment evaluates the effect of three usability mechanisms on a web application. The examined usability mechanisms are: Abort Operation, Progress Feedback and Preferences. It was established that the three usability mechanisms require the inclusion of additional components because they affect system functionality and not just its interface [6].

Although the field of SE experimentation has already established the importance of running replications and analysing families of experiments for improving the quality of evidence, there are still very few studies that have applied this concept. There is one study comparing three requirements elicitation methods [21], whose aim is to help developers select the best method. Other studies analyse model-driven development (MDD) in terms of final software quality [20] and maintainability [22]. MDD is a software engineering approach applying models and model technologies to raise the level of abstraction at which developers create and evolve software, with the goal of both simplifying and formalizing the various activities and tasks that comprise the software life cycle. We also found studies that define a framework or evaluate MDD tool usability [20], [23], which could be used to conduct families of experiments.

One family of experiments evaluated whether the use of test-driven development (TDD) improves software product quality [25]. The family is composed of 12 separate experiments and aims to improve the accuracy and generalizability of the results. TDD is an agile software development approach that uses testing-coding cycles. The study evaluates whether the characteristics of the experiments affect the results of TDD performance in terms of quality.

Finally, the only study that we have found that uses families of experiments to evaluate usability-related aspects plans a family of experiments to empirically evaluate a web usability evaluation process (WUEP) proposed by the authors [26] within the framework of MDD use. There were 64 participants in the family of experiments, including PhD and MS computer science students. The objective of the experiments was to evaluate the participants' effectiveness, efficiency, perceived ease of use and perceived satisfaction when using WUEP compared to heuristic evaluation (HE), which is an inspection method widely used in industry.

The inclusion of additional components in a software product implies increased development time and costs. Studies addressing the inclusion of usability mechanisms have concluded that it is more costly (in terms of time or money) to implement some mechanisms than others [5], [14], [47]. As there is a difference in implementation costs, the results of the experimental evaluation of the real effect of the mechanisms on final system usability can help to prioritize mechanism implementation.

As a result of the analysis of the related work, we have found only one paper using families of experiments to evaluate application usability according to HCI recommendations. This study uses a family of experiments to improve the accuracy of the results on the implementation of specific usability mechanisms and their effect on the final usability of a web application. The results should provide SE with criteria to be able to evaluate and make more reliable decisions on the usability functionality that it is more feasible to implement within a specific system or web application.

## 3. BASELINE EXPERIMENT

This section reports the definition, design and settings of the baseline experiment. The details were published in a previous paper [14]. Two strict replications were conducted, which, together with the baseline experiment, constitute a family of three experiments.

### 3.1. Goal, research questions and hypotheses

The research goal of this experiment is to evaluate the impact of three usability mechanisms, ABR, PFB and PRF in a web application. The research question (RQ) is: Does the adoption of usability mechanisms have an impact on application usability?

The research question is further divided into three specific research questions:
- RQ1: Does the adoption of the ABR usability mechanism have an impact on application usability?
- RQ2: Does the adoption of the PFB usability mechanism have an impact on application usability?
- RQ3: Does the adoption of the PRF usability mechanism have an impact on application usability?

The aim of the specific research questions is to check whether or not the adoption of ABR, PFB and PRF, respectively, improves the application usability in terms of efficiency, effectiveness and user satisfaction

The null hypothesis governing these three specific research questions is: H.1.x.0: There is no significant difference in user EFFICIENCY | EFFECTIVENESS | SATISFACTION with or without the adoption of the usability mechanism. This hypothesis is broken down into three specific null hypotheses, one for each quality characteristic (where x represents 1. Efficiency, 2. Effectiveness and 3. Satisfaction). For RQ1, the three hypotheses are:

- H.1.1.0: There is no difference in EFFICIENCY with or without the adoption of ABR.
- H.1.2.0: There is no difference in EFFECTIVENESS with or without the adoption of ABR.
- H.1.3.0: There is no difference in SATISFACTION with or without the adoption of ABR.

The three null hypotheses for RQ2 and RQ3 are formulated similarly.

### 3.2. Factors and response variables

The factor or independent variable [36] defined for the family of experiments is the usability mechanism with two levels: adopted and not adopted. *Adopted* means that a specified usability mechanism is adopted during task performance. *Not adopted* indicates that a specified usability mechanism is not adopted during task performance.

The baseline experiment aimed to evaluate the effect of the usability mechanism through the response variables: efficiency, effectiveness and user satisfaction. According to ISO/IEC 25010 [4], efficiency refers to resources expended by users to correctly and completely achieve specific goals, effectiveness is the degree to which users correctly and completely achieve specific goals and satisfaction is the degree to which users' needs are satisfied by using a product or system in a specified context of use.

In the following, we describe the metrics used for each response variable —efficiency, effectiveness and satisfaction—:

- Efficiency is measured according to two metrics:
  a) Speed: time measured in seconds taken by a subject to complete the task [49]. The clocked time represents the time taken by the subject to perform the task and, if necessary, to reread the instructions during task performance. Efficiency measured as user speed can be represented by:

  $$Ef_{speed} = \frac{StopTime_{milliseconds} - StartTime_{milliseconds}}{1000}$$

  b) Interactivity: number of clicks made by a subject to complete the task [50], [51]. We count separate clicks, where a double click is classed as two separate clicks. Efficiency measured as user interactivity can be represented by:

  $$Ef_{interactiveness} = count(separateClicks)$$

- Effectiveness: percentage task completion by a subject [48]. Effectiveness can be represented by:

$$Effectiveness = \frac{Number\ of\ successfully\ completed\ subtasks}{Total\ number\ of\ subtasks\ undertaken} * 100\%$$

- Satisfaction: mean value of the responses to the post-task questionnaire questions. The questionnaire responses are ordinal values on a Likert scale (1 = Strongly disagree to 5 = Strongly agree) [52]. There are two satisfaction questions per mechanism. Satisfaction can be represented by:

$$s = \frac{questionValue_1 + questionValue_2}{2}$$

### 3.3. Context and experimental subjects

The baseline experiment was conducted in two contexts: academic setting and non-academic setting [14]. The experiment was executed in each context at different non-overlapping time periods. The experimental subjects were not computer science specialists. The experiment had a total of 168 participants: 88 from the academic setting and 80 from the non-academic setting.

The biggest concentration of participants spanned two main age groups: 18–30 years (61%) and 31–40 years (26%). All the subjects had to perform the tasks set as part of the experiment. Participation was voluntary and, in the case of students, required the consent of the institutional authorities.

### 3.4. Experimental design

The family of experiments uses a between-subjects design with orthogonal array [53], [54]. Each experimental subject was placed in a single group and sequentially performed randomly assigned tasks to interact with all three (adopted or non-adopted) usability mechanisms. Accordingly, a single value is output for each experimental subject with respect to each response variable for use in the statistical analysis.

Our design is composed of a treatment matrix, a mechanism exposure order matrix and a group assignment matrix. Table 1 shows the treatment matrix describing which usability mechanisms will and will not be adopted. The zeros denote a non-adopted usability mechanism, whereas the ones stand for the adopted mechanism. For example, when a subject is assigned treatment A, he or she will have to perform the ABR and PFB tasks without access to the usability mechanism and the PRF task with the enabled usability mechanism.

TABLE 1
TREATMENT MATRIX

| Treatment | ABR | PFB | PRF |
|---|---|---|---|
| A | 0 | 0 | 1 |
| B | 0 | 1 | 0 |
| C | 1 | 0 | 0 |
| D | 1 | 1 | 1 |

Table 2 shows the order of exposure for each factor. This matrix establishes all the possible task performance sequences for each factor (without repetitions).

Finally, each row of the treatment matrix is combined with each row of the exposure order matrix to produce 24 groups (group assignment matrix).

TABLE 2
MECHANISM ORDER EXPOSURE MATRIX

| Order | Task 1 | Task 2 | Task 3 |
|-------|--------|--------|--------|
| O1 | ABR | PFB | PRF |
| O2 | ABR | PRF | PFB |
| O3 | PFB | PRF | ABR |
| O4 | PFB | ABR | PRF |
| O5 | PRF | ABR | PFB |
| O6 | PRF | PFB | ABR |

### 3.5. Instrumentation and tasks

The family of experiments uses a web application system: an online shop called QuickStore [55], [56]. The application and user interface design include automatic group and task allocation, as well as data collection. Each subject performs three tasks. The tasks are:

- Abort Operation: the subject applies a cancel operation to his or her shopping cart. Upon login, the user's shopping cart will already contain several items. The user has to go to his or her shopping cart and modify data (for example, increase the number of any of the items, enter a promotional code, etc.) and then cancel the operation. If the usability mechanism has been adopted, the user will have a quick cancel option and will merely have to confirm the cancellation of all the pending changes. If the usability mechanism has not been adopted, the user will have to manually undo each change made since the start of the task one by one.
- Progress Feedback: the subject has to search for a specified item and add this item to the shopping cart. The subject starts the task from the QuickStore application home page, running a search using his or her preferred criteria, for example, item name. If the search is successful, he or she merely has to press the Add to Shopping Cart button. If the usability mechanism is enabled, a progress bar will be displayed while the search is running telling the user that the action is being executed and a message will be displayed at the end of the search specifying the number of items found. If the usability mechanism has not been adopted, the user will not be informed during the search that the action is ongoing.
- Preferences: this task is divided into two parts. The user will perform first the basic task and then the fictitious task. Basic Task: the subject should customize the application user interface. The font size of the original interface is small and not very legible. If the usability mechanism has been adopted, the user can customize some shop features to his or her liking. On the other hand, if the mechanism has not been adopted, the user cannot modify the application interface appearance. Fictitious Task: the user is asked to search for information on the time limit for returns of purchased items provided by the application. If the subject has modified the system interface, he or she can easily find the link to the required information. However, if the user was not able or decided not to modify the application interface appearance, it will be very hard for him or her to find the required information.

### 3.6. Operation

The baseline experiment was conducted over a five-month period from March to July 2016. Over the first four months, the experiment was executed within academia at the Autonomous University of Asunción using the distance education platform (e-campus[1]). Each professor published the experiment link on his or her course e-campus. Over the last month (July), the experiment was conducted outside academia. Subjects were informed that participation was voluntary. The students that agreed to participate were encouraged to do their best to perform the tasks, although it was an optional challenge that had no bearing on their learning outcomes.

At the time of experiment execution, the subjects were not familiar with the aim of the study or with the research hypotheses. Apart from the link [55] that each participant was to use to log in and start the evaluation, we did not provide any additional material. Originally, data for a total of 182 subjects were collected. However, data for 14 subjects were removed because they did not correctly complete the tasks. Finally, 168 valid data remained for the statistical analysis and results interpretation.

## 4. REPLICATIONS

The baseline experiment concluded that the impact of a mechanism may depend on other factors and vary depending on the context. Therefore, it was necessary to conduct further experiments to gather more evidence and confirm the results.

This section describes the replications conducted, highlighting the similarities and differences to the baseline experiment. The two replications are real experiments conforming to a between-subjects design. They have the same goal, research questions, hypotheses and instrumentation. Replication 1 measures two of the response variables: efficiency and satisfaction. The experimental data on effectiveness are missing. Replication 2 measures all three response variables: effectiveness, efficiency and satisfaction. The replications differ as to the experimental subjects.

To describe each replication of the original empirical research, we apply the guidelines defined for reporting experimental replications proposed by Carver [57]. To analyse the family of experiments, we apply Steps 1 to 3 of the guidelines recommended by Santos et al. [19]:

- Step 1: Describe the participants
- Step 2: Analyse individual replications
- Step 3: Aggregate the results.

### 4.1. Replication 1

The experimental subjects of Replication 1 are students from Rodeira Secondary School in Galicia, Spain, who volunteered to participate in the experiment with the consent of their teachers. This replication was conducted in August 2016. To rule out the learning effect, no informative or practice sessions were held beforehand. All the subjects completed a familiarity questionnaire. The details of the sample are as follows:

- The sample was composed of 100 subjects, of which 43 were male and 57 female.
- With regard to age, 85 of subjects were in the 16 to 18 age group, 10 were members of the 18 to 30 age group, and 5 were aged over 30.
- The subjects connect to the Internet at home. Some also use the Web at work or elsewhere. The primary uses are for entertainment and education.
- With respect to subjects' online shopping habits, most participants had never purchased anything over the Internet (37%), whereas 32% shopped online occasionally, 17% rarely, 9% almost always and only 5% always. As with our baseline experiment, this is an advantage as most subjects are not acquainted with the application domain and are therefore more sensitive to system usability.

---
[1] http://e.uaa.edu.py/

### 4.2. Replication 2

This experiment was run in an academic setting with first-year students of accountancy, law, sport sciences and health sciences at the Autonomous University of Asunción, all of whom participated on a voluntary basis. This replication was conducted over a two-month period from November to December 2016. The experimental subjects were not computer science specialists. This experimental constraint underpinned the idea that subjects with little or no computer expertise can use the system and appreciate the benefits of usability. Additionally, it rules out the influence of computer-literate users who may be familiar with this type of applications.

Like the baseline experiment and Replication 1, no informative session was held beforehand, again to rule out the learning effect. Participants were given an overview of the application, introducing the structure of the experiment to assure that they were able to successfully perform the tasks. The results of the familiarity questionnaire completed by subjects before the start of the experiment were as follows:

- The final sample included 99 subjects, of which 45 were male and 54, female.
- Most participants were aged from 18 to 30 years (89%), except nine that were members of the 31 to 40 age group (9%) and two within the 41 to 50 age group (2%).
- The experimental subjects are regular Internet users. They connect to the Web at home (73%), at work (19%) and to a lesser extent elsewhere (8%).
- Most participants had never shopped online (66%), whereas 15% rarely, 15% occasionally, and 4% more often (always or almost always) shopped online.

Table 3 provides a detailed summary of the subjects that participated in the family of experiments, specifying the differences between the baseline experiment, and Table 4 describes the profile of the researchers that participated in each experiment from design to results analysis.

### 4.3. Analysis of replications

We analyse the replications following the same procedure enacted in the baseline experiment [14]. Briefly, we divide the analysis into three parts, one for each usability mechanism: ABR, PFB and PRF. In each part, we evaluate the impact of usability: clicks, times, percentage task completion

TABLE 3
SUMMARY OF SUBJECTS

|  | Baseline experiment | Strict Rep1 | Strict Rep2 |
| --- | --- | --- | --- |
| Subjects type | Academic and non-academic | Academic | Academic |
| Number of participants | 168 | 100 | 99 |
| Men | 76 | 43 | 45 |
| Women | 92 | 57 | 54 |
| Age range with the largest number of participants | 18-30 | 16-18 | 18-30 |

*Rep means Replication: both terms are used indistinctly hereinafter.*

TABLE 4
SUMMARY OF EXPERIMENTERS

| Experimenters | Baseline experiment | Strict Rep1 | Strict Rep2 |
| --- | --- | --- | --- |
| Designer | Teachers from UPM-UAM-UNA | Teachers from UPM-UAM-UNA | Teachers from UPM-UAM-UNA |
| Monitor | Teachers from UNA | Teachers and student from UAM | Teachers and student from UAM |
| Measurer | Teachers from UNA | Teachers and student from UAM | Teachers and student from UAM |
| Analyst | Teachers from UPM-UAM-UNA | Teachers from UPM-UAM-UNA | Teachers from UPM-UAM-UNA |

and satisfaction. Two groups were compared: a group in which the usability mechanism was adopted and a group in which the mechanism was not adopted. We provide the violin and box plots and evaluate the statistical significance (p-value).

We use the Mann-Whitney U test [58] to evaluate statistical significance. Note that the Mann-Whitney U test is a scale-free statistical test and can assess the statistical significance of all response variables irrespective of the data type (continuous, discrete, ordinal, etc.).

We have not removed any outliers because they are regarded as legitimate experiment values.

### 4.3.1. Abort Operation

Table 5 shows the summary statistics for each response variable distribution (depending on whether the ABR usability mechanism is or is not adopted) of all the replications. The respective violin and box plots are show in Fig. 1. The line between the two boxes connects the means.

**Efficiency**. As Fig. 1 shows, the subjects using the system with adopted ABR appear to be more efficient in terms of clicks and time. Table 5 shows that the difference is statistically significant (in terms of clicks and time) for Replication 2.

**Effectiveness**. The result shows that there is a considerable difference between adopted and non-adopted ABR in Replication 2 (Fig. 1). Table 5 shows that this difference is significant and appears to be larger when ABR is adopted. There are no data on effectiveness for Replication 1.

**Satisfaction**. Fig. 1 indicates that there is an observable increase in the mean satisfaction across replications: the subjects appear to be more satisfied when ABR is adopted. The difference in user satisfaction is found to be statistically significant in all replications.

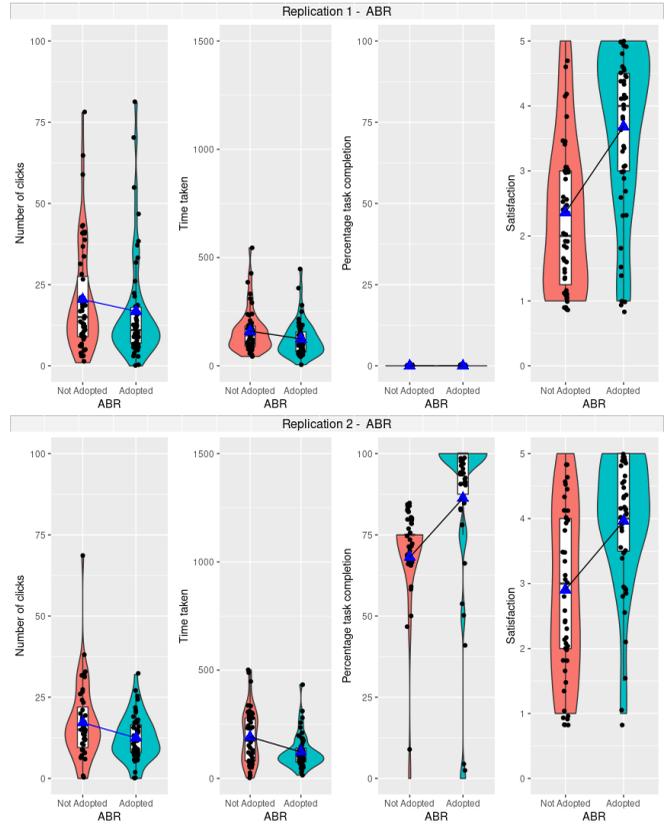

Fig. 1. Violin plots for the number of clicks, elapsed time, percentage task completion and satisfaction with the ABR usability mechanism: Rep1 and Rep2

TABLE 5
SUMMARY STATISTICS AND STATISTICAL SIGNIFICANCE ASSESSMENT FOR ABR: REP1 AND REP2

| Response Variable | Rep. | Group | Mean | Median | SD | p-value |
|---|---|---|---|---|---|---|
| Click | Rep1 | Not adopted | 20.41 | 15 | 16.76 | 0.11 |
| | | Adopted | 16.80 | 11 | 16.85 | |
| | Rep2 | Not adopted | 17.13 | 15 | 12.12 | 0.04 * |
| | | Adopted | 12.49 | 12 | 6.93 | |
| Time | Rep1 | Not adopted | 160.01 | 141.072 | 100 | 0.03 * |
| | | Adopted | 124.73 | 103.2 | 83.28 | |
| | Rep2 | Not adopted | 191.61 | 174.02 | 123.54 | 0.006 * |
| | | Adopted | 122.37 | 103.64 | 77.17 | |
| Percentage | Rep1 | Not adopted | - | - | - | - |
| | | Adopted | - | - | - | |
| | Rep2 | Not adopted | 68.09 | 75 | 19.30 | <0.001 * |
| | | Adopted | 86.27 | 100 | 28.86 | |
| Satisfaction | Rep1 | Not adopted | 2.36 | 2 | 1.22 | <0.001 * |
| | | Adopted | 3.68 | 4 | 1.27 | |
| | Rep2 | Not adopted | 2.90 | 3 | 1.30 | <0.001 * |
| | | Adopted | 3.96 | 4 | 1.02 | |

Comparing the results of the replications with the baseline experiment for ABR, we find that Replications 1 and 2 return similar results to the baseline experiment. The adoption of ABR improves efficiency (speed), effectiveness and user satisfaction. However, the adoption of ABR does not appear to improve user efficiency in terms of interactivity.

The analysis for PFB and PRF is described in the Appendix.

## 5. ANALYSIS APPROACH

Due to the heterogeneity of the resulting impacts in the three experiments considering the three usability mechanisms, it would be premature to draw conclusions based on the separate results of each experiment. Besides, aggregation procedures would mitigate the threat to the generalization and reliability of the results of the individual experiments [14], [19]. Note that rather than reproduce the published baseline experiment results, our aim is to pool together the different experiments in order to understand the effect of the usability mechanism in a broader setting [18].

In our case, the family of experiments is composed of three experiments: (a) Baseline Experiment, with 168 subjects; (b) Replication 1, with 100 subjects; and (c) Replication 2, with 99 subjects.

We divide the analysis of the family of experiments into three different parts, one per usability mechanism: Abort Operation, Progress Feedback and Preferences. We assess four response variables according to each usability mechanism: CLICK (i.e., number of clicks), ELAPSED_TIME (i.e., time taken to perform the task), PERCENTAGETASK (i.e., the percentage task completion) and VALUE (i.e., the satisfaction score on a 1-to-5 Likert scale). The PERCENTAGETASK response variable was not measured in Replication 1.

We follow an identical analysis procedure for each usability mechanism (i.e., within each part):
- We provide a profile plot showing the average score of the subjects for each response variable divided by the adoption/non-adoption of the usability mechanism across the replications. We make preliminary observations with respect to the differences in the results across the experiments.
- Following the conventions used in medicine to analyse groups of interrelated experiments, we fitted fixed-effects linear regression models with the main factor TREATMENT and EXPERIMENT to analyse the data [59], [60]. We chose linear regression over meta-analysis of effect sizes [46], as: (1) access to the raw data is guaranteed within the family, and (2) all the replications have identical response variable operationalizations. We fitted a fixed-effects linear regression instead of a random-effects model (i.e., linear mixed model with EXPERIMENT as a random factor and TREATMENT as a random effect) because: (1) experiment operationalizations are identical, and (2) populations are similar across the replications. Two relevant assumptions need to be met within fixed-effects linear models: the normality assumption and the homoscedasticity assumption (i.e., the equality of variances across the treatment groups [60]). The normality assumption is tenable due to the relatively large sample size achieved at the family level (i.e., sample size in the hundreds [61], [62]). With regard to the homoscedasticity assumption, we fitted generalized least squares models [63] accommodating different variances across treatment groups and experiments (i.e., allowing for heteroscedasticity) to assess the robustness of the linear regression results. As the linear regression and generalized least square results were similar, we chose to interpret the statistical significance and practical significance of results using the most parsimonious model (i.e., the linear regression model).
- We assessed the statistical significance of results according to the p-value of the TREATMENT estimate. We assessed the practical significance of the results according to: (1) the sign of the TREATMENT estimate, and (2) the magnitude of the TREATMENT estimate with respect to the intercept term (i.e., the non-adopted condition in the baseline experiment, since the non-adopted condition in the baseline experiment is taken as the reference class in all the fitted fixed-effects regressions). If the control conditions differ markedly across the replications (and, thus, the estimate of the control condition in baseline experiment is uninformative for assessing the magnitude of the TREATMENT estimate), we assess the magnitude of the TREATMENT estimate considering the control estimates of the other experiments also.
- To ease the integration of results in future meta-analyses [46], we provide Cohen's d effect sizes [64], alongside their interpretation (i.e., small, medium, large) and their corresponding variances for all the pairwise comparisons made (i.e., the adoption/non-adoption of each usability mechanism for each response variable) for all the experiments. We used the R package effsize [65] to compute the effect sizes and their respective variances.

Throughout this section, we analyse the data of each usability mechanism one by one (i.e., abort operation, progress feedback and preferences).

### 5.1. Abort Operation Analysis

Fig. 2 shows the profile plot for CLICK, ELAPSED_TIME, PERCENTAGETASK and VALUE by adoption/non-adoption of the Abort Operation usability mechanism across all experiments.

As shown in Fig. 2, the averages are consistent across the replications: the adoption of the abort mechanism decreases (1) the number of clicks and (2) the elapsed time across all the experiments, and increases (1) the percentage of task completion and (2) subject satisfaction.

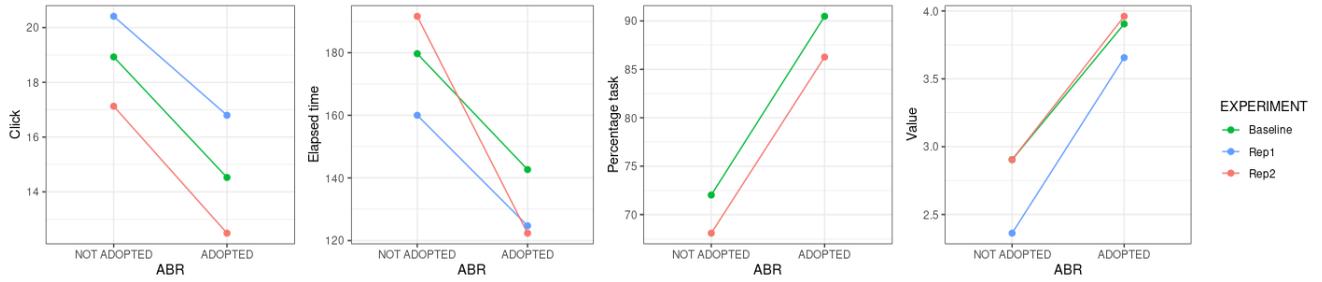

Fig. 2. Profile-plot for ABR

Table 6 shows the results of the linear regression models fitted to analyse the adoption or non-adoption of Abort Operation across experiments.

TABLE 6
LINEAR REGRESSION COEFFICIENTS FOR ABR

| Coefficient | Click | Time | Percentage | Satisfaction |
|---|---|---|---|---|
| Intercept | 18.85 (1.21)*** | 183.76 (10.99)*** | 72.07 (2.05)*** | 2.85 (0.11)*** |
| Adopted | -4.25 (1.36)** | -45.16 (12.33)*** | 18.36 (2.56)*** | 1.10 (0.12)*** |
| Experiment==Rep1 | 1.87 (1.64) | -18.90 (14.90) | | -0.38 (0.15)* |
| Experiment==Rep2 | -1.93 (1.65) | -4.68 (14.99) | -4.07 (2.65) | 0.03 (0.15) |

*Significance levels: \*\*\* (p<0.001), \*\* (p<0.01), \* (p<0.05), '.' (p<0.1).*

As Table 6 shows, the Abort Operation usability mechanism appears to have a remarkable impact on the number of clicks (i.e., a drop in the number of clicks of around 23% (i.e., 4.25/18.85) with respect to the intercept: the average score calculated for the non-adopted abort operation condition in the baseline experiment). This drop appears to be larger for elapsed time (i.e., a drop of around 24% in time). We also find an increase of around 25% in percentage task completion, and a larger increase in satisfaction (i.e., an increase of almost 39%). Thus, overall, the adoption or non-adoption of the Abort Operation usability mechanism appears to have a major impact on system usability.

Table 7 shows Cohen's d effect sizes, interpretations (i.e., small, medium, large), and respective variances for each replication.

TABLE 7
REPLICATION EFFECT SIZES FOR ABR

| Response variable | Experiment | d | vi | Interpretation |
|---|---|---|---|---|
| Click | Baseline | -0.3634 | 0.0242 | small |
| | Rep1 | -0.2151 | 0.0402 | small |
| | Rep2 | -0.4748 | 0.0420 | small |
| Time | Baseline | -0.2672 | 0.0240 | small |
| | Rep1 | -0.3827 | 0.0407 | small |
| | Rep2 | -0.6786 | 0.0432 | medium |
| Percentage | Baseline | 1.0102 | 0.0268 | large |
| | Rep1 | - | - | - |
| | Rep2 | 0.7352 | 0.0436 | medium |
| Satisfaction | Baseline | 0.8543 | 0.0260 | large |
| | Rep1 | 1.0398 | 0.0459 | large |
| | Rep2 | 0.9053 | 0.0451 | large |

### 5.2. Progress Feedback Analysis

Fig. 3 shows the profile plot for CLICK, ELAPSED_TIME, PERCENTAGETASK and VALUE by adoption/non-adoption of the Progress Feedback usability mechanism across all experiments.

As Fig. 3 shows, the sign of the effects appears to be consistent across all experiments, except for percentage task completion (where the baseline average appears to be unchanged irrespective of whether the mechanism is or is not adopted). Of all the experiment participants, Replication 2 subjects appear to experience the largest drop in number of clicks, elapsed time, and percentage task completion. Besides, with respect to Replication 1 and Baseline Experiment subjects, there is a smaller increase in satisfaction among Replication 2 participants when the mechanism is adopted.

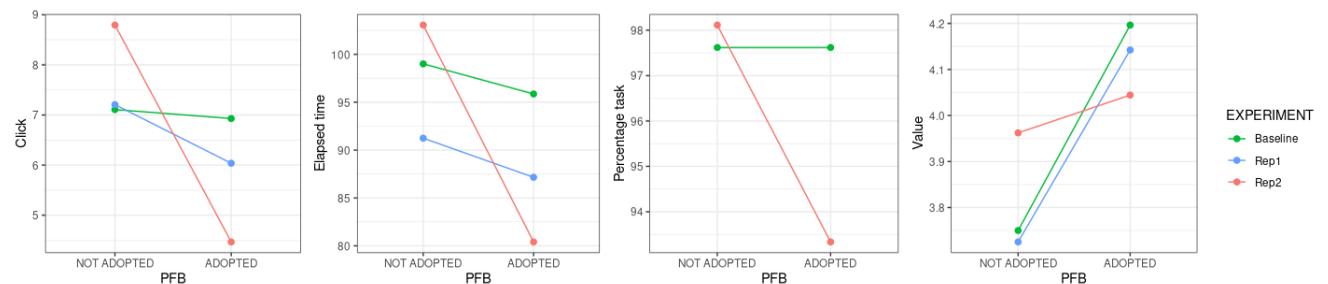

Fig. 3. Profile-plot for PFB

Table 8 shows the results of the linear regression models fitted to analyse the adoption or non-adoption of the Progress Feedback mechanism across experiments.

TABLE 8
LINEAR REGRESSION COEFFICIENTS FOR PFB

| Coefficient | Click | Time | Percentage | Satisfaction |
|---|---|---|---|---|
| Intercept | 9.86 (1.10)*** | 111.29 (8.80)*** | 30.74 (4.49)*** | 2.05 (0.11)*** |
| Adopted | -3.15 (1.08)** | -18.79 (8.61)* | 42.10 (5.59)*** | 1.44 (0.11)*** |
| Experiment==Rep1 | 0.09 (1.23) | 5.86 (9.79) | | -0.11 (0.12) |
| Experiment==Rep2 | -0.09 (1.58) | 2.39 (12.64) | -1.19 (5.79) | 0.38 (0.15)* |

*Significance levels: \*\*\* (p<0.001), \*\* (p<0.01), \* (p<0.05), '.' (p<0.1).*

As shown in Table 8, the Progress Feedback usability mechanism appears to affect the response variables to a smaller extent than the other usability mechanisms. Specifically, the number of clicks and elapsed time are reduced by around only 18% and 7%, respectively, and the percentage task completion by a negligible amount (i.e., around 2%). Besides, the satisfaction scores do not appear to increase much either (just 10%). In view of these findings, the adoption or non-adoption of the Progress Feedback mechanism does not appear to have much of an impact on system usability. Interactions may be in operation, and, therefore, moderators should be identified in order to explain the heterogeneity of results.

Table 9 shows Cohen's d effect sizes, interpretations (i.e., small, medium, large), and respective variances for each replication.

TABLE 9
REPLICATION EFFECT SIZES FOR PFB

| Response variable | Experiment | d | vi | Interpretation |
|---|---|---|---|---|
| Click | Baseline | -0.0250 | 0.0238 | small |
| | Rep1 | -0.1910 | 0.0150 | small |
| | Rep2 | -0.5272 | 0.0425 | medium |
| Time | Baseline | -0.0328 | 0.0238 | small |
| | Rep1 | -0.0511 | 0.0149 | small |
| | Rep2 | -0.2445 | 0.0414 | small |
| Percentage | Baseline | -0.0000 | 0.0238 | small |
| | Rep1 | - | - | - |
| | Rep2 | -0.2408 | 0.0414 | small |
| Satisfaction | Baseline | 0.4831 | 0.0245 | small |
| | Rep1 | 0.4762 | 0.0154 | small |
| | Rep2 | 0.0794 | 0.0411 | small |

### 5.3. Preferences Analysis

Fig. 4 shows the profile plot for CLICK, ELAPSED_TIME, PERCENTAGETASK and VALUE by adoption/non-adoption of the Preferences usability mechanism across all experiments.

As shown in Fig. 4, the direction of the effects is consistent across the experiments: while the Preferences mechanism decreases the number of clicks and the elapsed time, it increases percentage task completion and the user satisfaction.

Table 10 shows the results of the linear regression models fitted to analyse the adoption or non-adoption of the Preferences mechanism across experiments.

As Table 10 shows, the Preferences usability mechanism appears to have a considerable effect on the number of clicks (i.e., leading to a drop in the number of clicks of around 32%), a smaller effect on the elapsed time (i.e., leading to a drop of around 17% in time), a larger effect on task completion (i.e., an increase of almost 137%) and satisfaction (i.e., an increase of around 70%). Therefore, the adoption or non-adoption of the Preferences usability mechanism appears to have a substantial impact on system usability, especially percentage task completion and user satisfaction.

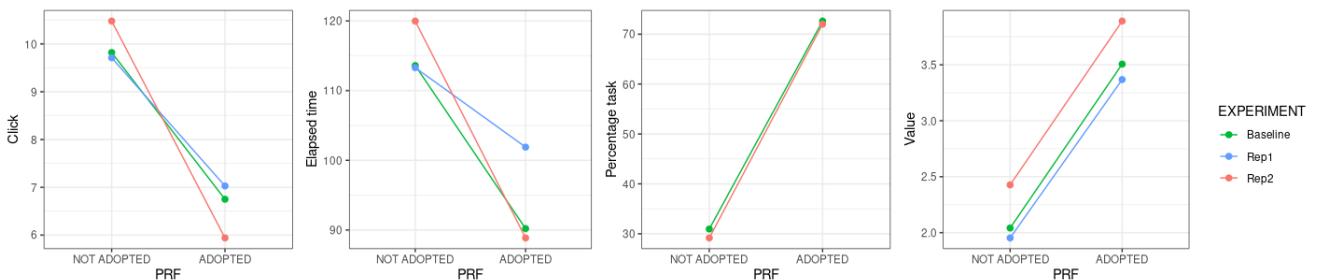

Fig. 4. Profile-plot for PRF

TABLE 10
LINEAR REGRESSION COEFFICIENTS FOR PRF

| Coefficient | Click | Time | Percentage | Satisfaction |
|---|---|---|---|---|
| Intercept | 7.73 (0.61)*** | 101.02 (7.74)*** | 98.50 (1.69)*** | 3.79 (0.08)*** |
| Adopted | -1.43 (0.60)* | -7.18 (7.58) | -1.75 (2.10) | 0.37 (0.08)*** |
| Experiment==Rep1 | -0.39 (0.68) | -8.20 (8.61) |  | -0.04 (0.09) |
| Experiment==Rep2 | -0.27 (0.88) | -5.07 (11.13) | -1.77 (2.18) | 0.04 (0.12) |

*Significance levels: \*\*\* (p<0.001), \*\* (p<0.01), \* (p<0.05), '.' (p<0.1).*

Table 11 shows Cohen's d effect sizes, interpretations (i.e., small, medium, large), and respective variances for each replication.

TABLE 11
REPLICATION EFFECT SIZES FOR PRF

| Response variable | Experiment | d | vi | Interpretation |
|---|---|---|---|---|
| Click | Baseline | -0.2292 | 0.0240 | small |
|  | Rep1 | -0.2162 | 0.0150 | small |
|  | Rep2 | -0.4136 | 0.0417 | small |
| Time | Baseline | -0.2391 | 0.0240 | small |
|  | Rep1 | -0.1184 | 0.0150 | small |
|  | Rep2 | -0.2805 | 0.0412 | small |
| Percentage | Baseline | 0.9119 | 0.0263 | large |
|  | Rep1 | - | - | - |
|  | Rep2 | 0.9385 | 0.0453 | large |
| Satisfaction | Baseline | 1.2363 | 0.0284 | large |
|  | Rep1 | 1.1333 | 0.0174 | large |
|  | Rep2 | 1.2307 | 0.0486 | large |

## 5.4. SUMMARY OF THE RESULTS

In this section, we discuss the quantitative results in response to the research questions. Table 12 summarizes the experiment results. The percentage values signify the ratio of the adopted factor (i.e., 4.25/18.85 for clicks) to the intercept, that is, the average score calculated for the non-adopted condition in the baseline experiment). The (-) sign means that the adoption of a mechanism has a negative effect on the response variable. The ∗ symbol denotes that the input is significant.

### 5.4.1. *RQ1 Abort Operation*

The adoption of Abort Operation has a favourable impact on efficiency, effectiveness and user satisfaction. Table 12 shows, for efficiency, an improvement of around 23% in user speed and interactivity. Effectiveness and user satisfaction are higher when this mechanism is adopted (25% and 39%, respectively). Additionally, the input of this mechanism is statistically significant for all response variables. *We conclude that the adoption of ABR improves efficiency, effectiveness and user satisfaction.*

### 5.4.2. *RQ2 Progress Feedback*

Table 12 shows that, on the one hand, user interactivity and satisfaction were slightly better (i.e., 18% and ~10%) for subjects that had access to the mechanism than for others that did not. In both cases, the input is statistically significant. On the other hand, the adoption of PFB does not play a key role in either effectiveness or efficiency in terms of user speed. *We conclude that the adoption of PFB does not improve efficiency, effectiveness and user satisfaction.*

### 5.4.3. *RQ3 Preferences*

The results shown in Table 12 confirm that PRF has a positive impact on all response variables. In particular, the adoption of PRF was a decisive factor for improving effectiveness and user satisfaction (137% and 70%, respectively). Also there is evidence that users are 17% faster (take less time) and interact less with the system (32%) (use fewer clicks) to perform the specified tasks. The input of this mechanism is statistically significant. *We conclude that the adoption of PFB improves user efficiency, effectiveness and satisfaction.*

## 6. VALIDITY THREATS

In this section, we discuss the threats with respect to the statistical conclusion validity, internal validity and external validity.

TABLE 12
SUMMARY OF THE EXPERIMENT RESULTS

| Usability Mechanism | Efficiency | | Effectiveness | Satisfaction |
|---|---|---|---|---|
|  | Clicks | Time | Percentage | Value |
| ABR | (-) 23% * | (-) 24% * | 25% * | 39% * |
| PFB | (-) 18% * | (-) 7.1% | (-) 1.8% | 9.8% * |
| PRF | (-) 32% * | (-) 17% * | 137% * | 70% * |

## 6.1. STATISTICAL CONCLUSION VALIDITY

The threats at the level of the family of experiments that are related to the conclusion validity appear when replicating the experiment and combining the results. We relied upon parametric statistical tests (i.e., LMM [66]) to analyse the data of our family of experiments. We ensured the robustness of the results that we provided by meta-analysing the data with the one-stage IPD model and an extra factor that accounts for the difference between results across experiments [19], [20]. In order to ensure the transparency of the results, the original data and statistical analyses carried out are provided in the supplementary material. All the supplementary material is also available at figshare (URL: https://doi.org/10.6084/m9.figshare.13148117).

## 6.2. INTERNAL VALIDITY

To increase internal validity, participants were not informed about the tasks that they were to perform beforehand. In the following, we discuss the five identified threats to internal validity and the actions taken to mitigate these threats.

The first two threats related to technological expertise and the order of task performance. With regard to technological expertise, although all the experiment participants are novices with regard to their level of experience with this type of experiments, they do not all have the same expertise regarding the activity to be performed. Besides the familiarity questionnaire revealed that a large percentage subjects were familiar with the use of web pages, although online shopping rate among subjects is low. As far as the order of task performance is concerned, there could be bias caused by the learning effect, as the tasks associated with each mechanism are performed sequentially.

To mitigate the above two threats, the subjects were randomly assigned to balanced groups. This randomization procedure is an experimental guarantee [67], as interferences may or may not occur irrespective of their impact. It is worthwhile making the effort to randomize experiments to offset any potential bias.

A third internal validity threat is low user experience, where there is a risk of users not making the effort it takes to understand the instructions, comprehend the procedure, etc. We overcome this threat by introducing the order as a design factor.

The fourth threat is related to the fact that subjects perform the usability test remotely, and it is not possible to interact with participants in real time. As a result, the participants could perform the experiment more than once, do things wrong or drop out of the experiment because they misunderstand the task instructions and do not have the chance to ask what to do when they are unsure. To try to mitigate this threat, we captured the IP address of each subject and an additional contact address (for example, telephone number, email address or chat ID). We used the IP address to exclude any subjects that performed the experiment more than once. We used the contact information to gather feedback from the subject.

Finally, there is a fifth internal validity threat related to motivation. Each participant will, foreseeably, react differently to the experiment, and subjects may perform poorly, especially if they are alternating experiment performance with other activities. This threat cannot be mitigated. Nonetheless, we interviewed subjects at random to find out if they suffered from fatigue, boredom or similar. The responses should be taken into account during the analysis and interpretation of the results to reduce the impact of this threat.

## 6.3. EXTERNAL VALIDITY

We identified two threats to external validity. The first threat is that experimental results cannot be generalized to all users. To prevent any potential bias caused by familiarity with the technology, the participants selected to participate in all three experiments are not computer scientists. Nevertheless, all the participants are regular Internet users. Additionally, the subjects are members of a sizeable user population group that tends not to use online shopping web applications. However, we can gather quite reliable empirical evidence about the impact of the usability mechanisms analysed at lay user level.

Another probable threat is the generalization to applications from other domains. This threat could be dealt with by executing the experiment in other application domains.

## 7. CONCLUSIONS

Having run a set of three experiments (the baseline experiment, and two replications), we conducted a meta-analysis applied to the family of experiments. To do this, we used the linear regression model at family level with reference to the baseline experiment.

The aggregate data of the family of experiments again suggest that the adoption of the three mechanisms improves system usability and does not undermine user performance. In particular, a decisive improvement is not always observed in the case of the efficiency response variable, although the adoption of the usability mechanism never detracts from user efficiency. In the case of Abort Operation, the improvement in user efficiency is conclusive. In the case of Progress Feedback, the difference between adoption and non-adoption of the mechanism is appreciable only for number of clicks. Finally, for Preferences, the difference between mechanism adoption and non-adoption is conclusive for number of clicks and quite large for time reduction.

In the case of the effectiveness response variable for Abort Operation and Preferences, the difference between adoption and non-adoption of each mechanism improves system usability conclusively. There is nosignificant improvement in the case of Progress Feedback.

For the satisfaction response variable, there is a significant difference between adoption and non-adoption for both Abort Operation and Preferences, which should therefore be taken into account, whereas the improvement is almost non-existent for the Progress Feedback mechanism.

Based on the meta-analysis results, we can therefore conclude that adoption and non-adoption of the Abort Operation and Preferences mechanisms appear to have a major impact on system usability with respect to both user efficiency, effectiveness and satisfaction. Progress Feedback appears to affect the response variables less, and, ultimately, has a negligible impact on system usability.

Most of the values are statistically significant. The effect of the values that are not significant is so small that it is not worth pursuing further research because, thanks to the high number of experimental subjects, the evidence that we have gathered from the replications would overrule the results of other experiments [18].

In sum, the family of experiments endorses the result of the baseline experiment and further ratifies the finding that Progress Feedback does not lead to appreciable improvements in user performance (at least with respect to the task implemented within the study domain).

This research is another step forward in the empirical analysis of usability from the user viewpoint. Additionally, the statistically significant findings are reason enough to further pursue experimentation by changing the design of the task to create more complex scenarios, modifying the experimental design by altering the instrumentation for application in a different domain from online shopping, and researching other usability mechanisms from the perspective of users applying the same quality attributes to check their impact on web environments.

Finally, families of experiments are becoming increasingly important in SE, generating evidence underpinning the evolution of knowledge on the impact of the usability recommendations implemented through usability mechanisms in web environments.

## ACKNOWLEDGMENT

Work funded by FEDER/Spanish Ministry of Science and Innovation-Research State Agency (projects MASSIVE, RTI2018-095255-B-I00, and PGC2018-097265-B-I00) and the R&D programme of Madrid (project FORTE, P2018/TCS-4314).

## SUPPLEMENTARY MATERIAL

Available at Code Ocean (https://doi.org/10.24433/CO.1678000.v1)

1. **Analysis.R**: R code to analyze experiments (Summary statistics).
2. **Meta-analysis.R**: R code to analyze SE family of experiments.
3. **Baseline.xlsx**: Raw data of baseline experiment.
4. **Replication_1.xlsx**: Raw data of replication #1 of baseline experiment.
5. **Replication_2.xlsx**: Raw data of replication #2 of baseline experiment.

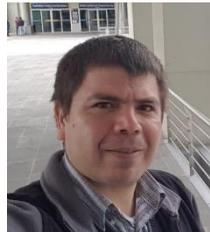

**Juan M. Ferreira** received his MSc in Software and Systems from the Technical University of Madrid in 2011 and his MSc in ICT majoring in Software Engineering from the National University of Asunción in 2018. He is currently assistant professor of software engineering at the National University of Asunción. His fields of interest are usability and experimental software engineering.

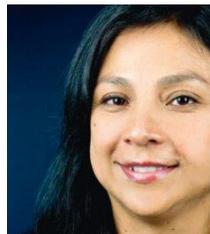

**Francy D. Rodríguez** received her PhD from the Technical University of Madrid (UPM) in 2015. She is currently associate professor of computer science at the University of Ávila. Her research interests include software development, design and programming patterns, and software usability.

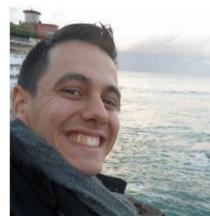

**Adrián Santos** received his MSc in Software and Systems and MSc in Software Project Management from the Technical University of Madrid, Spain, and his MSc in IT Auditing, Security and Government from the Autonomous University of Madrid, Spain. He received his PhD in Software Engineering from the University of Oulu, Finland. He is currently working as a software engineer in industry.

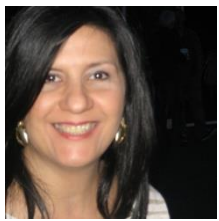

**Silvia T. Acuña** received her PhD from the Technical University of Madrid in 2002. She is currently associate professor of software engineering at the Autonomous University of Madrid's Computer Science Department. Her research interests include experimental software engineering, software usability, software process modelling and software team building. She co-authored A Software Process Model Handbook for Incorporating People's Capabilities (Springer, 2005), and edited Software Process Modelling (Springer, 2005) and New Trends in Software Process Modelling (World Scientific, 2006). She is deputy conference co-chair of the ICSE 2021 organizing committee. She is a member of the IEEE Computer Society and a member of the ACM.

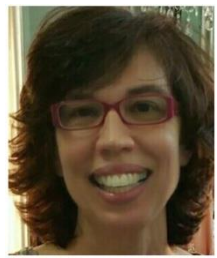

**Natalia Juristo** received her PhDfrom the Technical University of Madrid (UPM) in 1991. She is currently full professor of software engineering at UPM. She received a Finland Distinguished Professor Program (FiDiPro) professorship, starting in January 2013. Her main research interests include experimental software engineering, requirements, and testing.